# Analysis of a Planetary Scale Scientific Collaboration Dataset Reveals Novel Patterns

Soumya Banerjee [1,2,3]

[1]Harvard University
 Boston, USA

[2]Ronin Institute
 Montclair, USA

[3]Complex Biological Systems Alliance
 North Andover, USA

```
neel.soumya@gmail.com
```

**Abstract**

Scientific collaboration networks are an important component of scientific output and contribute significantly to expanding our knowledge and to the economy and gross domestic product of nations. Here we examine a dataset from the Mendeley scientific collaboration network. We analyze this data using a combination of machine learning techniques and dynamical models. We find interesting clusters of countries with different characteristics of collaboration. Some of these clusters are dominated by developed countries that have higher number of self connections compared with connections to other countries. Another cluster is dominated by impoverished nations that have mostly connections and collaborations with other countries but fewer self connections. We also propose a complex systems dynamical model that explains these characteristics. Our model explains how the scientific collaboration networks of impoverished and developing nations change over time. We also find interesting patterns in the behaviour of countries that may reflect past foreign policies and contemporary geopolitics. Our model and analysis gives insights and guidelines into how scientific development of developing countries can be guided. This is intimately related to fostering economic development of impoverished nations and creating a richer and more prosperous society.

# 1 Introduction

Scientific collaboration networks are an important component of scientific output and contribute significantly to expanding our knowledge and contribute to gross domestic product. Collaboration networks have been modelled before [1, 2]. Here we examine a novel dataset for the Mendeley collaboration network [3]: we find novel clusters of countries with different characteristics of collaboration. We propose a dynamical model that explains these characteristics. We also find interesting patterns in the behaviour of the countries that may reflect foreign policies and contemporary geopolitics.

# 2 Collaboration Data

We use a planetary scale collaboration data freely available for download from Mendeley Labs [3]. Analysis of this data with k-means clustering reveals the following patterns:

a) We plot the percentage of external connections that each country has vs. the distinct number of countries each country is connected with. The plot shown below (Fig. 1) has a distinctive shape. Each point on the plot is a distinct country.

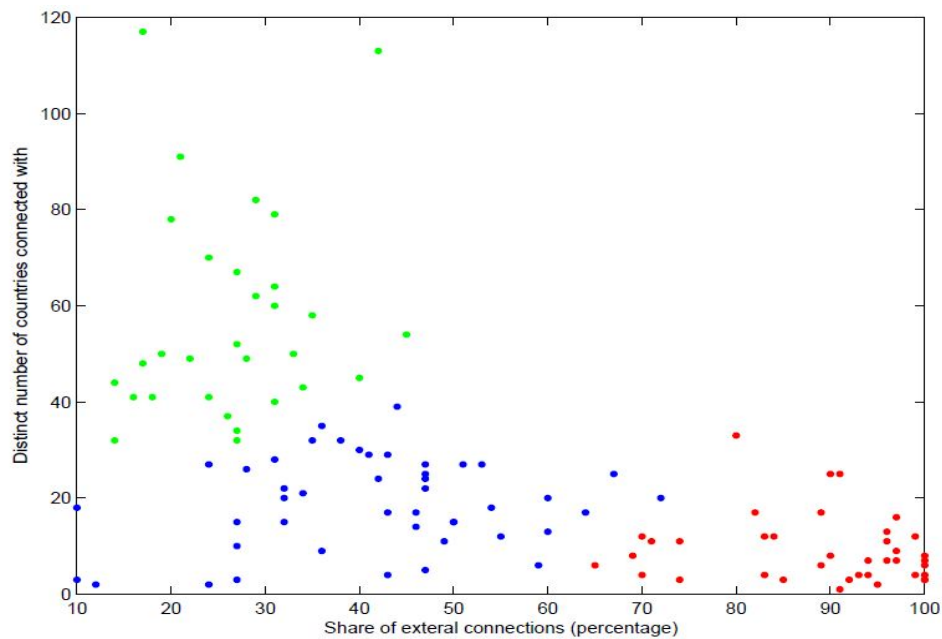

**Fig. 1.** Plot of the percentage of external connections that each country has vs. the distinct number of countries each country is connected with. Clustering is done with k-means and shows three distinct clusters.

b) There are 3 clusters corresponding to 3 quadrants (shown in three colours in Fig. 1); it is interesting to note that there is no country in the 4th quadrant (top right: the UK is close to the 4[th] quadrant).

c) There are no countries with high share and high distinct number of countries connected with. Countries with a larger share of foreign connections and less distinct countries connected with are less developed nations.

d) The country with the highest number of distinct number of countries connected with and largest share of external connections is the UK (top rightmost in the plot).

d) The country with the lowest number of distinct number of countries connected with and lowest share of external connections is Algeria (bottom leftmost in the plot).

In order to explain these patterns, we propose a dynamical model for how the links are actually formed; for example poor countries might preferentially seek out collaborations with richer countries or richer countries might help disadvantaged nations (e.g. Liberia has 100% external connections). We simulate the time evolution of connections. Our dynamical model correctly predicts the broad observations of scientific connections of countries.

## 3     Mathematical Model

Our dynamical model has two compartments: developing countries ($x$) and developed countries ($y$). Developing countries grow their scientific expertise and self-connections at a rate proportional to $\alpha*x$ (self-growth) and also by interacting with developed nations at a rate proportional to $\beta*x*y$. Developed nations ($y$) are assumed to have reached equilibrium of growth. The model, represented as differential equations is shown below:

$$\frac{dx}{dt} = \alpha x + \beta xy$$

$$\frac{dy}{dt} = 0$$

Shown below (Fig. 2) is a simulation of how scientific connections of developing countries evolve over time for a chosen set of parameters.

For the same set of parameters we also simulated how the percentage of foreign connections should vary (over time) with the number of self-connections (Fig. 3). This plot accurately recapitulates the patterns in the empirical data (Fig. 1). It shows there are only 3 quadrants. It is consistent with and effectively recapitulates empirical data.

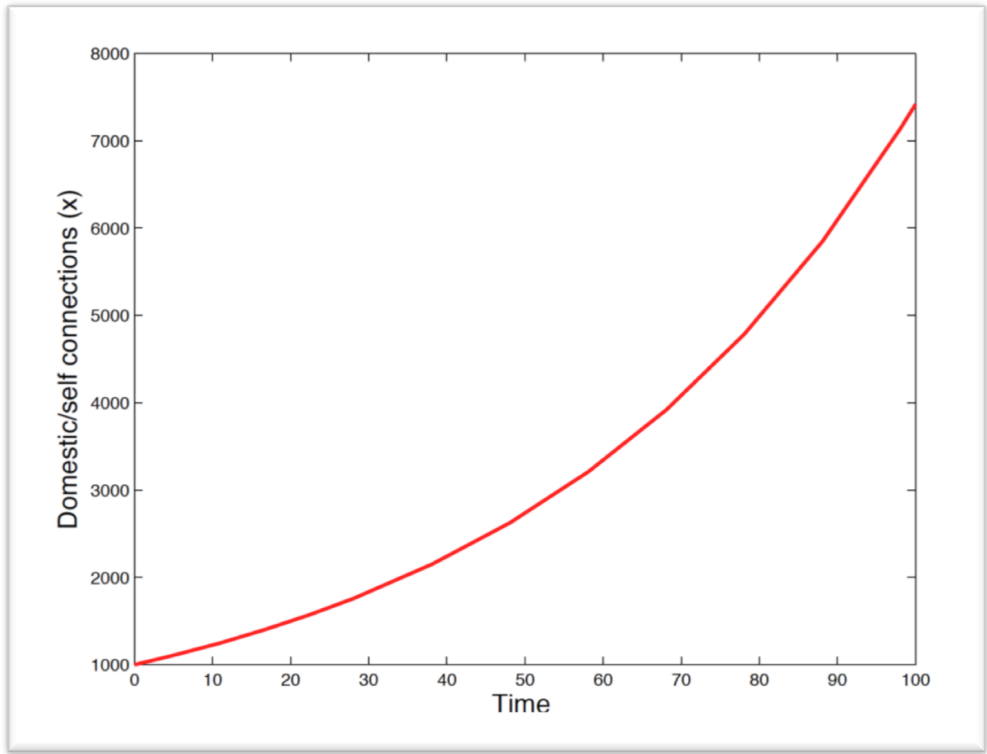

`Fig. 2.` Simulated plot of how scientific connections of developing countries evolve over time.

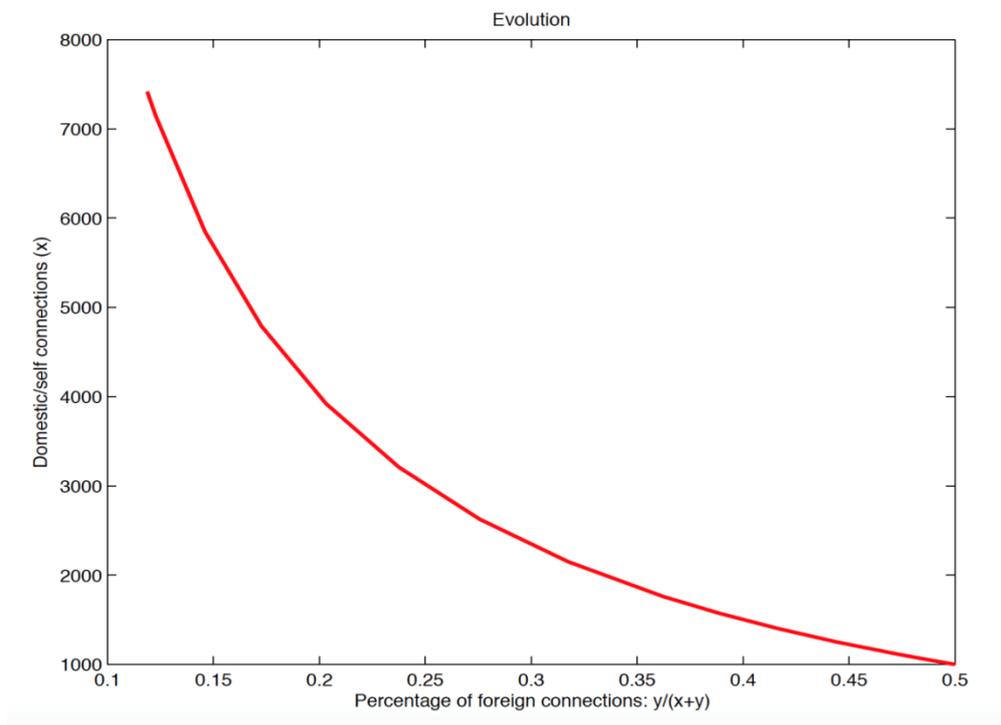

Fig. 3. Simulated plot of self connections vs. percentage of foreign connections.

## 4    Data and Statistical Methodology

All data is freely available from Mendeley Labs [3]. The table of the complete collaboration network is available for download [3]. The data was curated by Mendeley and a connection is defined as co-membership of two researchers in a Mendeley group. Further details of data pre-processing are available online [3].

All statistical analysis was conducted in MATLAB.

## 5    Discussion

We find novel patterns in the planetary scale collaboration dataset (Fig. 1). The table of the complete collaboration network is available for download [3]. We note some interesting patterns in the data:

1) The UK has a very high percentage of foreign connections; this maybe because of its colonial past.

2) Less developed or poorer countries usually have a very high percentage of foreign connections.

3) Iran has a large number of foreign connections; it is interesting that this is so despite foreign sanctions imposed against it.

4) The largest or richest countries have more distinct foreign connections. This may suggest that the rest of the world wants to collaborate and form connections with others; this could also be driven by interest in issues relevant to poorer countries like tropical diseases, socio-economic research into poverty and archaeological research in countries in Africa.

5) India has more foreign connections than China.

6) The US has a lower percentage of foreign connections but in absolute numbers it has the highest number of connections.

7) The highest percentage of foreign connections is usually occupied by very poor countries (presumably they are trying to build capability in science and technology by collaboration), e.g. Liberia has 100% foreign connections.

8) Among the developed nations, the UK has the largest percentage of foreign connections (probably reflecting its colonial heritage)

9) Some countries like El Salvador have a low percentage of foreign connections (this could be as a result of the fact that the country was embroiled in civil war for a long time). The policy implications are that it shows a need for targeted relief at starting active science and research programs in these nations.

10) South Korea, Japan, Taiwan, Japan, Germany have a low percentage of foreign connections (they have a large number of absolute connections and it is possible that they have invested very heavily in their own science programs). This is interesting since these are highly developed countries.

11) Liberia has 100% external connections: this suggests that more efforts need to be taken to develop its own scientific infrastructure.

12) Cuba has very few connections (this is again possibly due to sanctions imposed against it).

In summary we propose a dynamical model that explains current trends of clustering in scientific collaboration network. Taken together our data and models show signatures of past and contemporary foreign policies and geopolitics on the scientific collaboration landscape.

Our analysis also points to the power and generality of dynamical systems in modelling diverse sociological, biological and technological systems [4-15].

Our model and analysis gives insights and guidelines into how scientific development of developing countries can be guided. This is intimately related to fostering economic development of impoverished nations and creating a richer and more prosperous society.

## 6   Scientific Validation

This paper has been unanimously validated in a collaborative review mode with the following reviewers:

1) Maria Jose Jorente, Universidade Estadual Paulista

2) Mariana Cantisani, Federal University of Paraíba

## 7   References

1. The structure of scientific collaboration networks, Newman, Mark EJ, Proceedings of the National Academy of Sciences, 2001
2. World citation and collaboration networks: uncovering the role of geography in science, Pan, Raj Kumar and Kaski, Kimmo and Fortunato, Santo, Scientific reports, 2012
3. Mendeley Labs collaboration map, URL: http://labs.mendeley.com/collab-map/, Last accessed June 2015
4. Competitive dynamics between criminals and law enforcement explains the super-linear scaling of crime in cities. Banerjee S, Hentenryck PV, Cebrian M. Palgrave Communications 1, 2015.
5. Soumya Banerjee and Melanie Moses. Scale invariance of immune system response rates and times: perspectives on immune system architecture and implications for artificial immune systems. Swarm Intelligence, 2010
6. Soumya Banerjee, Scaling in the Immune System, PhD Thesis, University of New Mexico, USA, 2013
7. Liu, Peng and Calderon, Abram and Konstantinidis, Georgios and Hou, Jian and Voss, Stephanie and Chen, Xi and Li, Fu and Banerjee, Soumya and Hoffmann, Jan Erik and Theiss, Christiane and Dehmelt, Leif and Wu, Yao Wen, A bioorthogonal small-molecule switch system for controlling protein function in cells. Angewandte Chemie, 2014

8. Melanie Moses and Soumya Banerjee, Biologically Inspired Design Principles for Scalable, Robust, Adaptive, Decentralized Search and Automated Response (RADAR), Proceedings of the 2011 IEEE Conference on Artificial Life, 30-37, 2011.
9. Soumya Banerjee and Melanie Moses, Modular RADAR: An immune system inspired search and response strategy for distributed systems. In: E. Hart et al. (eds) Artificial Immune Systems, 9th International Conference, ICARIS, 2010, Lecture Notes in Computer Science, Springer Verlag: Berlin, Germany, vol 6209, pp 116–129, 2010.
10. Balch, Curt and Arias-Pulido, Hugo and Banerjee, Soumya and Lancaster, Alex K and Clark, Kevin B and Perilstein, Michael and Hawkins, Brian and Rhodes, John and Sliz, Piotr and Wilkins, Jon and Chittenden, Thomas. Science and technology consortia in US biomedical research: A paradigm shift in response to unsustainable academic growth. BioEssays.37(2),119-122,2015.
11. Soumya Banerjee and Joshua Hecker. A Multi-Agent System Approach to Load-Balancing and Resource Allocation for Distributed Computing, arXiv preprint arXiv:1509.06420, 2015
12. Soumya Banerjee and Melanie Moses, Immune System Inspired Strategies for Distributed Systems. arXiv preprint arXiv:1008.2799, 2010.
13. Soumya Banerjee and Melanie Moses, A hybrid agent based and differential equation model of body size effects on pathogen replication and immune system response. In: P.S. Andrews et al. (eds) Artificial Immune Systems, 8th International Conference, ICARIS 2009, Lecture Notes in Computer Science, Springer Verlag, Berlin: Germany, vol 5666, pp 14–18, 2009.
14. Soumya Banerjee, Drew Levin, Melanie Moses, Fred Koster and Stephanie Forrest. The value of inflammatory signals in adaptive immune responses. In: Lio, Pietro et al. (eds.) Artificial Immune Systems, 10th International Conference, ICARIS, Lecture Notes in Computer Science, Springer Verlag: Berlin, Germany, vol 6825, pp 1–14, 2011.
15. Soumya Banerjee. An Immune System Inspired Approach to Automated Program Verification. arXiv preprint arXiv:0905.2649. 2009.